\documentclass[prl,aps,twocolumn, showpacs,preprintnumbers,amsmath,amssymb]{revtex4}
\usepackage{graphicx}% Include figure files
\usepackage{dcolumn}% Align table columns on decimal point
\usepackage{bm}% bold math

\begin{document}

\title{Does Optical Anisotropy Lead to Negative Refraction at an Interface?}

\author{Ivan Biaggio}

\affiliation{
Center for Optical Technologies and  Department of Physics, \\
Lehigh University, Bethlehem, PA 18015
}

\received{\today}

\begin{abstract} This is a comment inspired by recently published
results [Y. Zhang et al., Phys. Rev. Lett. \textbf{91}, 157404 (2003)]
that introduced the name ``amphoteric refraction'' for the fact that at
the interface with an optically anisotropic material there can be a
range of incidence angles for which the  component of the Poynting
vector parallel to the interface changes sign upon refraction. The
latter effect is a well-known consequence of optical anisotropy, but it
was described as a new negative refraction phenomenon that can be
put in the same class as the negative refraction observed at an
interface with a left-handed material with negative refractive index. 
\end{abstract} 

\maketitle

In a recent letter (``Total Negative Refraction in Real Crystals
for Ballistic Electrons and Light'') \cite{Zhang03}, Zhang et al.\ studied light
propagation through the plane boundary between two different
orientations of the same birefringent crystal (``twin boundary''), with
the two optic axes parallel to the plane of incidence but tilted by the
same amount in  opposite directions with respect to the boundary. They
pointed out the range of incidence angles for which the components of
the incident and refracted Poynting vectors parallel to the interface
have a different sign. They called this effect ``amphoteric
refraction'', claiming that it was ``unusual'' and a case of negative
refraction.

\begin{figure}[b] \includegraphics[width=7cm]{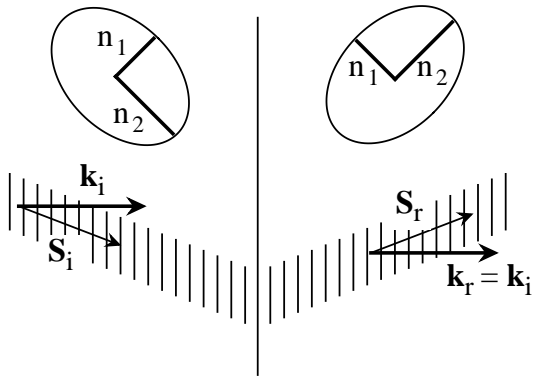} 
\caption{A twin boundary between two different orientations of an optically
anisotropic material (anisotropy ratio $n_2/n_1 = 1.5$). The indicatrix
is tilted by $-45^\circ$ and by $+45^\circ$ with respect to the
boundary. An optical wave that is polarized in the plane of the figure
is propagating with a wavevector ${\bf k}_i$ perpendicular to the interface. The
angle between wavevector and Poynting vector ${\bf S}_i$   is $\sim 21^\circ$ and
the latter has two different orientations in the two materials.}
\label{fig0} 
\end{figure}
 
However, the refraction effect described in Ref.~\onlinecite{Zhang03} is
simply an expression of the difference between Poynting vector and
wavevector in an electromagnetic wave  \cite{Born59}. It is misleading to put
it in the same class as the negative refraction effects that affect the
\textit{wavevector} of the wave, as  in the presence of a negative
refractive index \cite{Veselago68,Pendry00}. What has been described as a
new ``amphoteric'' refraction phenomenon in Ref.~\onlinecite{Zhang03} always
occurs at the interface with any anisotropic material when the main axes
of the indicatrix are not perpendicular to the interface, and leads to
such well-known effects as double-refraction in calcite and conical
refraction \cite{Born59}. A sketch of the interface in question is given
in Fig.~\ref{fig0}

\begin{figure}[t]
\includegraphics[width=8.5cm]{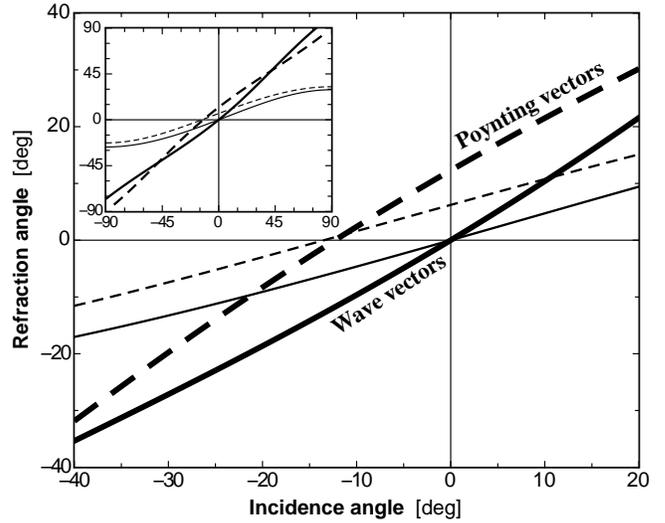}
\caption{ Refraction  vs.\ incidence angles in the same representation as 
in  Fig.\ 4 of Ref.~\onlinecite{Zhang03} and calculated
using the same material parameters. 
The dashed and solid curves refer to Poynting vectors and wavevectors, respectively. 
The thick curves are for the twin boundary, the thin curves are the result when the
 crystal on the incidence side is replaced by air. The inset shows
the full range of angles.}
\label{fig1}
\end{figure}

Fig.~\ref{fig1} plots the incidence and refraction angles in the same
way used in Ref.~\onlinecite{Zhang03}, with its data and theory given by the  thick dashed curves. 
The
solid curves give the corresponding wavevector angles, i.e.\ the incidence
and refraction angles used in Snell's law. The longitudinal and the
transverse components of the wavevectors of incident and refractive wave, and the corresponding
incident and refracted angles, always have the same sign, the
usual refraction effect expected for positive refractive indices.  The
fact that the dashed curves describing the Poynting vectors do not go
through the origin --- called ``amphoteric refraction'' in
Ref.~\onlinecite{Zhang03} --- is a normal effect in anisotropic materials which is
also seen at an air/crystal interface (thinner curves). Expressions for calculating 
refraction and reflectivity in
this case have been given, e.g., in Ref.~\onlinecite{stamnes77}.

These are  basic facts of wave propagation in anisotropic crystals, but
it is important to point this out in reference to the  claim of Ref.~\onlinecite{Zhang03}
of having obtained  negative refraction in the sense of
Ref.~\onlinecite{Pendry00}  ``despite all components of $\boldsymbol{\epsilon}$ 
and $\boldsymbol{\mu}$ 
being positive'' \cite{Zhang03} when what they showed is just the usual
effect of anisotropic (positive) refractive indices.

I think that in order to allow future discussions that focus on the
physics it highlights, the expression ``negative refraction''
should be used with some care. Simply applying the label to all effects
that cause the lateral component of the Poynting vector to switch sign
at an interface, thus including standard double-refraction effects, is a
generalization that does not appear to be very helpful in order to connect to the underlying
physical phenomena. It  seems more logical to apply
the ``negative refraction'' label in a more restrictive way, only to
those cases where unusual effects are observed in terms of the
wavefronts and wavevectors that are used in Snell's refraction law.

Since wave propagation is
described in terms of phase-changes and wavevectors --- which we have
seen behave normally at this interface ---  the discussion of
experimental opportunities and possible applications of the refraction
at the twin-boundary in Ref.~\onlinecite{Zhang03} should also be
re-examined.

As a final note, it must be pointed out that the present comment
should not overshadow another result derived in
Ref.~\onlinecite{Zhang03},
namely that the  reflection from the symmetric twin boundary is
exactly zero for all angles of incidence and despite the presence, at non-vanishing incidence angles, of a
finite refraction!  \cite{Liu04}

\textbf{Note:} Please see Refs.\ \onlinecite{Bliokh03} and\ \onlinecite{Yau03} for previous publications
that support the present comment, and Ref.\ \onlinecite{Ye04} for a
complementary, more general discussion of this topic.

\end{document}